# Dosimetric uncertainties related to the elasticity of bladder and rectal walls: adenocarcinoma of the prostate

Titre court: Dosimetric uncertainties and the adenocarcinoma of the prostate

Incertitudes dosimétriques relatives à l'élasticité de la paroi rectale et vésicale : Adenocarcinome de la prostate

Cyril Voyant[1,2], Katia Biffi [2], Delphine Leschi[2], Jérome Briançon[2], Céline Lantieri[2]

1 University of Corsica, CNRS UMR SPE 6134, (Campus Grimaldi, 20250 Corte), France

2 Hospital of Castelluccio, Radiotherapy Unit, BP 85, 20177 Ajaccio, France

Corresponding author: cyrilvoyant@hotmail.com  (Cyril Voyant, Tel:0495293666 and fax:0495293797)

**Abstract.**

*Background*

Radiotherapy is an important treatment of prostate cancer. During the sessions, bladder and rectal repletion is difficult to quantify and cannot be measured with a single and initial CT scan acquisition. Some methods, such as IGRT (Image Guided Radiation Therapy) and DGRT (Dose Guided Radiation Therapy), aim to compensate this missing information through periodic CT acquisitions. The aim is to adapt the patient's position, the beam configuration or the prescribed dose for a dosimetric compliance.

*Methods*

We evaluated organ motion (and repletion) for 54 patients after having computed the original ballistic on a new CT scan acquisition. A new delineation was done on the prostate, bladder and rectum to determine the new displacements and define organ doses mistakes (EUD, average dose and HDV).

*Results*

The new CT acquisitions confirmed that the bladder and rectal volumes were not constant during the sessions. Some cases showed that once validated treatment plan became unsuitable. A proposed solution is to correct the dosimetries when the bladder volume modifications are significant. The result is an improvement for the stability of the bladder doses, the D50 error is reduced by 25.3%, mean dose error by 5.1% and EUD error by 2.6%. For the rectum this method decreases the errors by only 1%. This process can reduce the risk of mismatch between the initial scan and the following treatment sessions..

*Conclusion*

For the proposed method, the cone beam CT is necessary to properly position the isocenter and to quantify the bladder and rectal volume variation and the deposited doses. The dosimetries are performed in the event that bladder (or rectum) volume modification limits are exceeded. To identify these limits, we have calculated that a tolerance of 10% for the EUD (compared to the initial value of the first dosimetry), this represents 11% of obsolete dosimetries for the bladder, and 4% for the rectum.

**Résumé.**

*Objectif*

La radiothérapie est un traitement important du cancer de la prostate. Durant les séances, les réplétions rectales et vésicales sont difficiles à quantifier, une unique tomodensitométrie initiale ne peut procurer cette information. Des méthodes comme l'IGRT (Image Guided Radiation Therapy) ou la DGRT (Dose Guided Radiation Therapy), ont pour but de compenser cette perte d'information, en utilisant des acquisitions CT périodiques. La finalité étant de proposer une radiothérapie adaptative et plus réelle.

*Méthode*

Nous avons évalués le mouvement (et la réplétion) des organes de 54 patients, après avoir calculé la balistique initiale sur une nouvelle tomodensitométrie. De nouvelles délinéations furent effectuées pour la prostate, la vessie et le rectum afin de déterminer les déplacements occasionnés, ainsi que les erreurs de doses générées sur les différents organes (EUD, dose moyenne and HDV).

*Résultats*

Les nouvelles acquisitions confirment que la vessie et le rectum ne sont pas de volume identique durant les séances. Dans certains cas la dosimétrie validée initialement n'est plus correcte et doit être retravaillée. Une solution proposée ici, est de ne s'attarder que sur les importantes modifications de volume vésical (5% des patients). Avec cette méthodologie, on constate une nette amélioration concernant la stabilité des doses vésicales, l'erreur sur le D50 est réduite de 25.3%, la dose moyenne de 5.1% et l'erreur sur l'EUD de 2.6%. Pour le rectum, cette méthode diminue toutes les erreurs de 1%. Ce processus tend à réduire le risque de non-respect de la dosimétrie validée, entre la simulation et les séances de radiothérapie.

*Conclusion*

La méthode proposée nécessite l'utilisation du cone beam CT pour bien positionner les volumes cibles mais aussi pour quantifier les variations de volume du rectum et de la vessie. Les distributions de dose sont retravaillées dés que les volumes vésicaux ou rectaux sont considérablement modifiés (limite fixée d'après la présente étude). Nous avons identifiés que pour une tolérance de 10% concernant les EUD (comparée à la valeur initiale établie sur le premier scanner), cela correspondait à 11% de dosimétrie obsolète pour la vessie, et 4% pour le rectum.

# 1 Introduction

Prostate cancer is the most frequent cancer affecting men over 50 years of age and second in terms of men mortality (25% of diagnosed cancers in France). Radiotherapy is an important treatment of prostate cancer. In some cases, it is the exclusive type of therapy. The current practice is dose escalation which limits the development of the tumour. The dosimetric limiting factor is the dose received by non-cancerous tissues, especially the organs at risk (OARs), such as the bladder, the rectum, femoral heads, the small intestine or the penis bulb [1-4]. In practice, rectum or bladder complications (late effects) often reduces the dose to the prostate. Dose escalation is really possible when the CTV motion during radiotherapy sessions (daily displacements or modifications) is controlled. The proposed study aims to, first, determine the mechanisms of organ interactions in the pelvic area and, second, estimate the consequences of these interactions on prostate treatment by radiation. In fact, for all the different methodologies used in prostate cancer treatment [5,6] (conformation, intensity modulation, arc-therapy, etc.), dosimetric planning uses data from the initial simulation alone [7]. To take into account organ repletion and internal morphology variation during the sessions, there are control methods such as IGRT [8,9] (image guided radiation therapy) and DGRT (dose guided radiation therapy) requiring periodic CT-acquisitions. The purpose is to adapt the patient's position, the beams and the prescription doses. These techniques decrease the errors due to patient positioning and ensure the proper location of the targets and OARs. In the proposed study, we chose to concentrate on the dosimetric consequences generated by bladder and rectum morphological changes. The first consequence is that initial skin markers (often tattoos for pelvic treatment) are no longer adapted and generate a loss of tumour control [10-12]. The same process can arrive to the tissues near the prostate. The physician and physicist validations (isodoses, organ dose, DVH…) related to the toxicity to the OARs cannot be adapted to the reality of treatment. The stochastic transformations which caused these patient setup errors are essentially related to the repletion of the bladder and rectum. These organs have a particular visco-elasticity that the radiotherapy team must quantify and control so as to increase the therapeutic dose (dose escalation). Elasticity is the quality of an object to be deformed and then resume its original shape when the stress applied disappears. These transformations are observable on pelvic organs, especially bladder or rectum muscles [13-15]. Firstly, we will describe the tools used and the clinical trials implemented. Then, we will reveal the statistical interaction between ballistic parameters and dosimetric values. Thirdly, we will consider the clinical solutions proposed to select the morphological modifications requiring a new dosimetry. Lastly, we will look at the impact of the daily CT acquisitions and the time required to control the dose to the organs.

## 2    Materials and methods

### 2.1    Methodology

The CT scan is a random snapshot of a changing patient. Organ position and repletion are fixed arbitrarily. One method of adaptive radiotherapy is the use of multiple scans (preferably one a day) to estimate the organ position and to adapt the beams. To protect the patient from excessive irradiation, we have only taken two CT scans per patient. One made during the treatment and another during the simulation, before the treatment began (necessary for treatment planning). With the new scanner, we evaluated the impact of time on the dosimetry, and verified that the ballistic was still appropriate. It is evident that, as has shown Van-Herk with analyses of organ motion from repeated CT [16], the intrinsic error of the dose on the irradiated volume has two distinct origins. One systematic and the other random, he suggests a necessary patient set-up margin which ensures a minimum CTV dose of 95% for 90% of the patient population, calculated as $2.5\Sigma + 0.7\sigma$ ($\Sigma$ is the standard deviation of the patient mean errors, and $\sigma$ the quadratic mean of the patient standard deviations). In the proposed study, the error due to movement is more "macroscopic" and the difference between the random ($\Sigma$) and the systematic ($\sigma$) term becomes negligible.

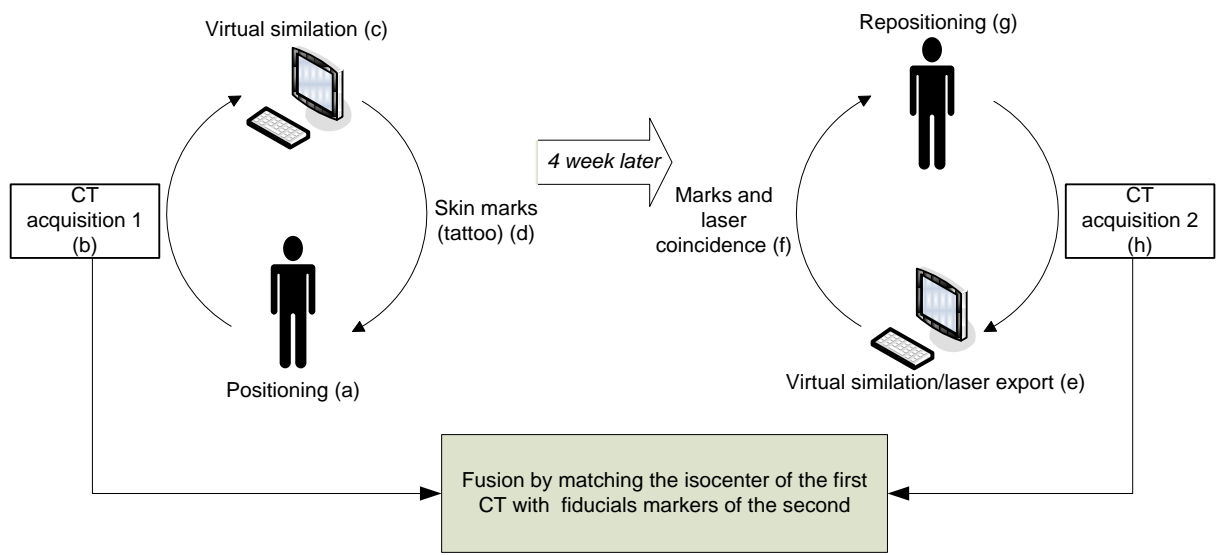

**Figure 1: Methodology for CT acquisitions**

As shown in Figure 1, evaluating organ motion and patient setup errors is done after computing the original ballistic on the new anatomical configuration; the dose to different organs can then be evaluated. The phase following the acquisition is the delineation on the new scan of the prostate, the bladder and the rectum by a well-trained person who has already delineated the first scan.

## 2.2  Patients eligibility and ballistics

From 2008 to 2009, 54 patients with organ-confined prostate cancer were prospectively enrolled on a conformational radiation treatment in the radiotherapy centre of Ajaccio, the CHD Castelluccio. Each patient was assigned a pre-treatment (virtual simulation) with a CT scan of 2-mm slice thickness, followed by a three-dimensional treatment plan. The initial CTV1 included the prostate and seminal vesicles and the second target volume (CTV2) included the prostate only. The PTV1 and PTV2 margins were done with CTV with a 1-cm uniform margin and 0.5-cm *in posterior*. The PTV-to-field-edge margin was 7-mm everywhere else, but 12-mm at the superior and inferior borders of the PTV. These margins take into account the dosimetric penumbra of the multileaf collimator and jaws. The selection criteria for this study were patients with non-operated prostate cancer at stage T2-T3aN0M0, with Gleason score less than or equal to 7 and with concentration of serum PSA less than or equal to 20 ng/ml. These patients were generally those for which the dose escalation treatment seemed most interesting. The majority of patients received a dose of 46 Gy on the PTV1 (4-beam-box-technique) and between 26 and 30 Gy boost on the PTV2 (generally 6 fields with different orientations). In the current study, normal tissues were delineated as follows: the rectal wall with a 0.5-cm wide ring (the extremities of the contouring were 1-cm below and above the CTV1 [17]); the bladder wall with a ring of 0.7-cm over the whole visible volume; the Treatment Planning System (TPS) used was Pinnacle® (Philips-AdacTM) equipped with the software version 8.0m with three-dimensional calculation with heterogeneity. Patients were in the supine position with the immobilization device Combifix® of SimMedTM. Patients were treated with a full bladder and an empty rectum. Only one patient was prescribed a laxative due to the presence of gas in the rectum. In addition, patients were not notified in advance of the day they received their second scan.

## 2.3 Comparison Criteria

### 2.3.1 Geometric critera

Organ movement induces dosimetric repercussions during the treatment. One or more beams can become unsuitable to the patient's anatomy, generating dosimetric errors on the target volumes and OARs. The moment at which these errors take place is unpredictable, and no prior action can help quantify this "bad" conformation. Many studies propose on-line or off-line solutions to take into account the morphological noise and properly target the PTV whatever the irradiation sessions [18,19,2021,22,23]. Rare are the solutions which can really be used on the influence of this noise on the OARs. Some authors believe that it is possible to calculate in real time the dose at any location on the patient [24,25]. This idea is not entirely realistic in a lot of radiotherapy unit. CT scans must be made frequently for all patients (maybe for all sessions) and the images acquired must be exported and merged in the planning system for organ delineation (~30 associated acquisitions). The next step must be the dose calculation with real volumes and real MU (Monitor Unit). It's only with the compilation of these plans (one plan per CBCT) that the true patient dosimetry can be obtained. In order for this process to be clinically possible, it is important to find shortcuts to avoid any unnecessary action. The idea is to determine a threshold value, below which no dosimetric correction is necessary, and above which, corrections are applied. Before making any dosimetric considerations, we will quantify the anatomic shifts and changes affecting the position of other organs. The purpose here is to find any interaction between the different pelvic volumes. We chose to consider the centroid (center of gravity) displacement and volume modification of bladder, rectum and prostate.

### 2.3.2 Dosimetric criteria

To finalize a dosimetric study, it's necessary to find and use the most relevant and objective parameters. In our case, we selected the most frequently used dosimetric criteria (in our unit) although more relevant ones could have been used instead. We chose the average dose, the Equivalent Uniform Dose (EUD) and dose to the volume for the OARs (D30 and D50 for the rectum and bladder respectively). Our lack of experience in 2008 concerning the parameter like V60, V70 and V74 for the rectum and the V60, V70 for the bladder (from "guide des procedures de 2007, SFRO" available on http://www.sfro.org/), forced us to did not use these tolerances. For the target volume, we only used the mean dose and D95. All TCP and NTCP values were inapplicable because a large dose difference is required to produce a significant difference between the probability values. It is possible to determine the EUD with some of the values available on the differential Histogram Dose Volume (HDV), and with the Pinnacle TPS (Equation 1). The parameters a and p (a= 1/p) being specific constants related to the tissue considered, $d_i$ the dose of each bin i (voxel i), the maximum of $d_i$ is noted $d_{max}$, $V_i$ the volume fraction associate to the voxel i and $vol_{ref}$ the volume of the organ in question. It is necessary to compute the effective volume of the organ, called *voleff,* to determine the EUD (Equation 2 and 3).

$$EUD = \left(\frac{1}{N}\sum (d_i^a)\right)^{1/a} = \left(\sum V_i . d_i^{1/p}\right)^p \qquad \text{Equ 1}$$

$$vol_{eff} = vol_{ref} . \sum V_i . \left(\frac{d_i}{d_{max}}\right)^{1/p} = \frac{vol_{ref}}{(d_{max})^{1/p}} . \sum V_i . d_i^{1/p}$$
$$= \frac{vol_{ref}}{(d_{max})^{1/p}} . EUD^{1/p}. \qquad \text{Equ 2}$$

$$EUD = d_{max} . \left(\frac{vol_{eff}}{vol_{ref}}\right)^p \qquad \text{Equ 3}$$

## 3   Results and interpretation

### 3.1   Centroid displacement

In this study, we found a very surprising setting: although the bladder volume was often completely different between the first and second scan, the repercussion on prostate displacement was often minimal. Table 1 shows the Mean Absolute Percentage Error (MAPE), which is defined in Equation 4; as the average of the absolute percentage deviation of the parameters in the two different scans ($V_{CT1}^i$ represents the volume of the considered organ measured on the CT1 and for the patient i).

$$MAPE = \sum_{i=1}^{N} \left| \frac{V_{CT1}^i - V_{CT2}^i}{V_{CT1}^i} \right| = \sum_{i=1}^{N} \left| 1 - 1 / (V_{CT1}^i / V_{CT2}^i) \right| \qquad \text{Equ 4}$$

It's shown that bladder volume was dramatically reduced by 50% on the second CT scan. This change didn't affect the prostate centroid which moved by only 0.5 cm; the displacement is isotropic in all three directions. The confidence interval for this organ is very low as the dispersion of values is so small. The question is, why was there such a large change in bladder volume between the two scans? We have worked on this variation and can conclude that the waiting time before and during the simulation must be reduced. From the 30[th] patient, we changed our methodology and acquired the CT scan maximum 1hr after the patient began to drink (results are not statistically significant but bladder volume ratio MAPE is equal to 51% for patients between the first and the 30[th] and 48% for others).

|  | C/C * | R/L* | A/P* | 3d | Vol (MAPE) |
|---|---|---|---|---|---|
| **Prostate** | 0,36 ± 0,09 | 0,24 ± 0,06 | 0,33 ± 0,08 | 0,54 | 13,06% |
| **Bladder** | 0,92 ± 0,31 | 0,27 ± 0,07 | 0,79 ± 0,16 | 1,24 | 49,63% |
| **Rectum** | 0,47 ± 0,10 | 0,28 ± 0,07 | 0,31 ± 0,06 | 0,63 | 17,69% |

**Table 1: absolute difference for the centroids (in cm), and the volumes between the first and second CT acquisition (*is the mean value and 95% confidence interval =mean±2.Std_dev.(54)$^{-1/2}$)**

Table 1 shows the volume change of the bladder which induced a centroid change in the C / C (cranial / caudal) and A / P (anterior / posterior) directions, but not in the L / R direction (left-right). We can see that for the three organs studied, the L / R orientation is not very conducive to the changes (0.2 - 0.3 cm). The results shown in this section are similar to those found in the literature [18]. The repletion phenomenon in the pelvis is very complex and to better understand its characteristics, it's necessary to test each organ correlation with other organ parameters, such as the displacement in the three directions or the organ volumes.

### 3.1.1 Bladder

Table 1 reveals a decrease in bladder volume at the second scan, as introduced in the previous section. In fact, the modification of the bladder's centroid follows a displacement axis upper anterior oblique.

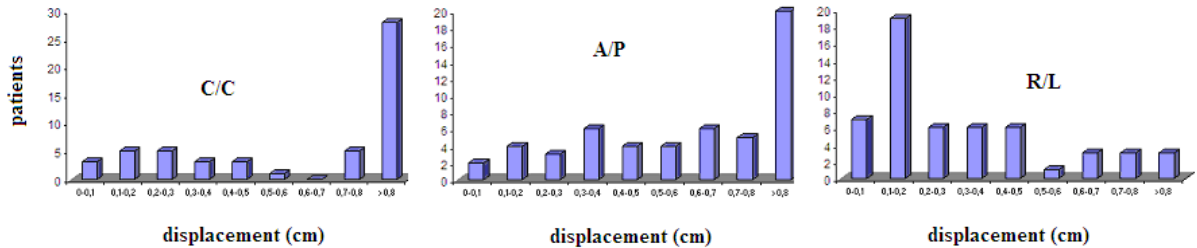

**Figure 2. Displacement of the bladder centroid**

The Figure 2 shows that for the bladder, the displacements are very important in the C/C and the A/P axis (> 0.8 cm for lots of patients). The correlation between the bladder volume ratio and the centroid offset in the A / P direction is equal to 0.28 (p = 0.06) and 0.58 for the C / C axis (p <0.001).

|  | Bladder | | Rectum | | Prostate | |
| --- | --- | --- | --- | --- | --- | --- |
|  | CC | p value | CC | p value | CC | p value |
| C/C Vs R/L | 0,08 | 0,55 | 0,02 | 0,89 | 0,004 | 0,98 |
| C/C Vs A/P | **0,33** | **0,03** | 0,23 | 0,11 | 0,06 | 0,66 |
| R/L Vs A/P | 0,26 | 0,08 | -0,14 | 0,32 | 0,1 | 0,47 |
| Age Vs 3d_disp | 0,23 | 0,12 | 0,11 | 0,42 | **0,29** | **0,05** |
| Vol Vs R/L | 0,24 | 0,11 | 0,03 | 0,79 | 0,03 | 0,78 |
| Vol Vs A/P | 0,28 | 0,06 | 0,06 | 0,69 | 0,04 | 0,79 |
| Vol Vs C/C | **0,58** | **<0,001** | **0,38** | **0,02** | 0,02 | 0,87 |
| Vol Vs age | -0,03 | 0,83 | -0,11 | 0,44 | -0,0004 | 0,99 |
| Vol Vs 3d_disp | **0,53** | **0,002** | 0,29 | 0,06 | 0,06 | 0,67 |

**Table 2: correlation coefficients (CC) for the bladder, rectum and prostate parameters**

Table 2 shows that a change in the A / P axis is often followed by one in the C / C axis (p = 0.03). These two parameters are linked, and given this characteristic, instructions were given to the manipulation team to regularly ask patients to come to radiotherapy with full bladders. A change in repletion follows a change in volume, which then incurs a centroid change in the C /C and the A /P directions. The bladder repletion compliance must ensure the stability of the bladder centroid. In this case the displacement of the isocenter will of about 0.3 cm in all directions, (equal to the shift observed in the R / L direction which is not correlated with organ volume).

### 3.1.2 Rectum

For the rectum, the displacement in the C / C axis (Fig. 3) is only an indication because the delineation is linked to the fixed limits of the CTV1 and CTV2.

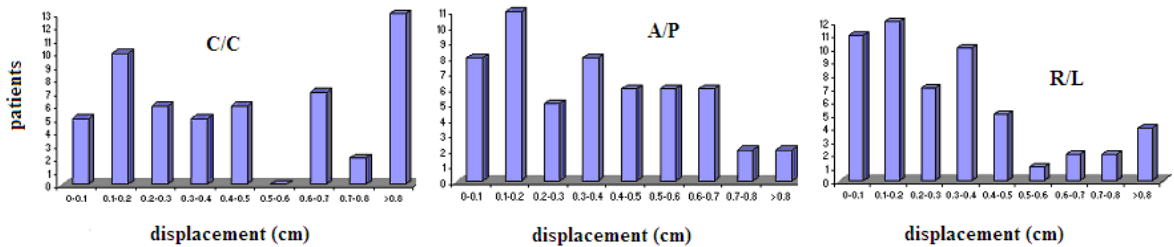

**Figure 3: Rectum centroid displacements**

Although the distributions aren't Gaussian, we realized that the averages for the R / L and A / P offsets are approximately 0.5 cm. Table 2 reveals a significant correlation between the volume of the rectum and the C / C rectum displacement (correlation coefficient CC = 0.38, $p < 0.02$).

### 3.1.3 Prostate

Margins applied to our CTV are 1 cm across and 0.5 cm posteriorly.

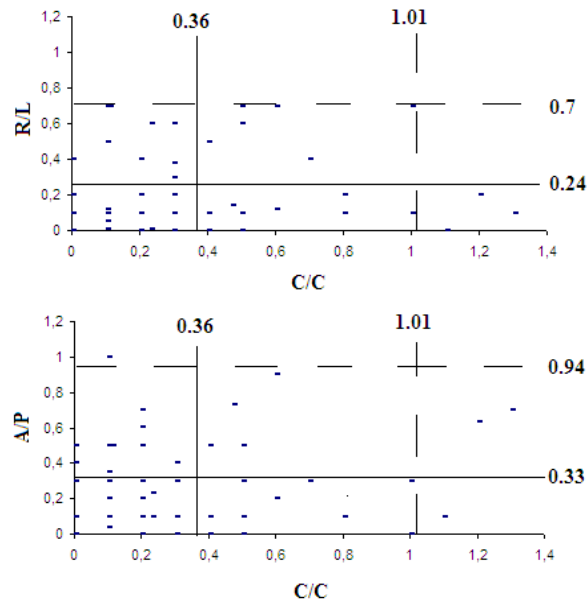

**Figure 4: Prostate centroid displacement (cm). The line is the average, and the dotted line the average + 2$\sigma$**

Fig. 4 shows that for many patients, the centroid offsets can exceed these values (~ 5%). It is important to understand the mechanisms to better control and predict an eventual dosimetric drift. The volume ratio is distributed around 1, and certainly must be due to the uncertainty of intra-operative delineation (white noise, Fig. 5).

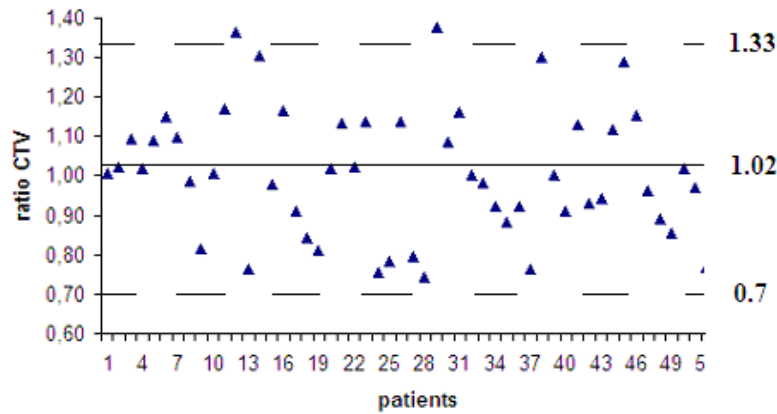

**Figure 5 : CTV volume ratio between the two scans, line is the average and dotted line the average +/- 2 σ (standard deviation)**

Table 2 does not allow to draw any conclusion because many of the correlations are not significant for the selected patient pool. There is a noticeable phenomenon related to the patients' age (Table 2), which affects the position of the isocenter between the two scans spaced in time (p = 0.05). It's difficult to determine the exact causes of this interaction, but it's possible that the sphincters are damaged and become fragile with age. Thus, bladder and rectum repletion cannot be maintained during the sessions and the prostate is then displaced. To test this hypothesis we will test the cross-interaction between volumes and centroids for the bladder, rectum and prostate.

### 3.1.4 Cross-interactions

The cross-interaction study allows to determine the most relevant factors among all organs. The results of these interactions are shown in Table 3.

| Prostate parameters | Bladder parameters | CC | p-value | Rectum parameters | CC | p-value |
|---|---|---|---|---|---|---|
| Vol | 3d_disp | -0,16 | 0,26 | 3d_disp | -0,12 | 0,42 |
| 3d | Vol | 0,08 | 0,60 | Vol | 0,16 | 0,27 |
| Vol | Vol | -0,11 | 0,43 | Vol | -0,28 | 0,07 |
| **3d_disp** | **3d_disp** | **0,44** | **0,01** | **3d_disp** | **0,70** | **<0,001** |
| **C/C** | **C/C** | **0,38** | **0,02** | **C/C** | **0,56** | **<0,001** |
| R/L | C/C | -0,03 | 0,81 | C/C | 0,06 | 0,69 |
| A/P | C/C | -0,01 | 0,95 | C/C | 0,28 | 0,07 |
| A/P | **A/P** | **0,35** | **0,03** | **A/P** | **0,58** | **<0,001** |
| C/C | A/P | 0,10 | 0,48 | A/P | 0,21 | 0,16 |
| R/L | A/P | 0,16 | 0,28 | A/P | -0,08 | 0,60 |
| R/L | **R/L** | **0,48** | **<0,001** | **R/L** | **0,77** | **<0,001** |
| A/P | R/L | 0,08 | 0,56 | R/L | -0,08 | 0,59 |
| C/C | R/L | 0,07 | 0,63 | R/L | 0,05 | 0,73 |
| A/P | Vol | -0,02 | 0,89 | Vol | 0,23 | 0,13 |
| R/L | Vol | 0,02 | 0,89 | Vol | 0,04 | 0,80 |
| C/C | Vol | 0,13 | 0,37 | Vol | 0,12 | 0,39 |

| **Rectum** | **bladder** | **CC** | **p-value** |
|---|---|---|---|
| 3d_disp | Vol | -0,09 | 0,53 |
| Vol | 3d | 0,09 | 0,53 |
| Vol | Vol | 0,06 | 0,69 |
| **3d_disp** | **3d_disp** | **0,44** | **0,01** |
| C/C | C/C | 0,29 | 0,06 |
| C/C | R/L | 0,20 | 0,18 |
| C/C | A/P | 0,23 | 0,13 |
| A/P | A/P | 0,09 | 0,53 |
| A/P | C/C | 0,19 | 0,21 |
| A/P | R/L | 0,04 | 0,79 |
| **R/L** | **R/L** | **0,42** | **0,01** |
| R/L | A/P | 0,27 | 0,08 |
| R/L | C/C | 0,01 | 0,92 |

Table 3: **cross interactions between prostate and bladder and rectum parameters (top), and cross correlation between bladder and rectum parameters (bottom)**

The analysis reveals that when the centre of the bladder and rectum is shifted, the isocentre of the prostate is consequently shifted. In the case of the bladder, p = 0.02 for a shift in the C / C axis, p = 0.03 for an A / P axis shift, and p < 0.001 for an R / L axis shift. In the case for the rectum, p < 0.001 for a shift in all three axes. However, due to our sample size and methodology, we failed to uncover a correlation such as the one existing between an A / P displacement of the prostate and the rectal volume, although [21] reveal a significant correlation equal to 0.62. This interaction is therefore masked by other phenomena. Maybe there are manipulation errors related to metal markers or tattoos, which are sometimes not well positioned or the CT-fusion is not well done. The next section focuses on the dosimetric aspect related to organ movement. The question is: is the dosimetric validation set by physicists and physicians at the initial time t0, still valid with pelvic organ changes during radiotherapy sessions?

## 3.2 Dosimetric modifications

Considering the previous sections about the organ motion, it seems necessary to use the CBCT in all radiotherapy session in order to target the PTV. This is time-consuming for technicians (dosimetrists, physicians and physicists ) as well as for LINAC. Is it possible to make real-time dosimetries with volumes changing on a daily basis? First, we need to consider that there is a threshold level on the first dosimetry, which is considered for the parameters studied in the previous section (bladder or rectum repletion). If the idea is to re-delineate all contours when patients have large repletion modifications, it would be necessary to quantify the change that is considered significant. A simple approximation would be to redo only 5% of the dosimetries, as we try to limit the amount of redundant work. A wide range of rectum and bladder volumes would be collected from the samples of 54 patients. A CT acquisition during the treatment (scanner or CBCT) describes an increase or decrease of repletion. If the change is out of the 95% interval, it is decided to redo the dosimetry. The use of CBCT is relevant to position the isocentre of the prostate, but also to identify the critical cases where repletion is not conforming to the initial scan. In case of doubt, the CT image dataset must be exported to TPS, merged to delineate the volume and applied the ballistic, then the dosimetry will need to be revalidated.

### 3.2.1 Elimination of bladder parameter Outliers

In this first part, only patients with an important change in bladder volume were considered. A simple approximation is made by considering the volume ratio distribution as a Gaussian distribution. We know it is wrong given the nature of the definition of this parameter (ratio). This means that 0.5 and 2 are symmetric (increase or decrease by a factor 2), which does not correspond to a Gaussian distribution. The assumption of normality is strongly compromised, but we accept this approximation knowingly. In the cohort of the 54 patients, the MAPE of bladder volume between the two CT acquisitions is computed from the bladder volume ratio of each patient (see Equation 4). The average bladder volume ratio is 1.27 between the volume of the first and the second scan, and the standard deviation is 0.63. It was decided to refuse patients whose ratio is above 2.53 (= 1.27 + 2x 0.63), which corresponds to two patients in our case.

|         | C/C  | R/L  | A/P  | 3d   | Vol (MAPE) |
|---------|------|------|------|------|------------|
| without | 0,92 | 0,27 | 0,79 | 1,24 | 49,63%     |
| with    | 0,94 | 0,25 | 0,74 | 1,22 | 48,17%     |

**Table 4: geometrical impact for bladder of the 95% bladder outliers methodology**

Table 4 shows the effect of the elimination of these outliers. The impact is not apparent on the centroid displacement, but the gain is 1.5% for the MAPE of bladder volume ratio.

|  | Ratio Rectum | | | Ratio Bladder | | | Ratio CTV2 | |
|---|---|---|---|---|---|---|---|---|
|  | D30 | Dmean | EUD | D50 | Dmean | EUD | D95 | Dmean |
| Average (without) | 1,016 | 1,015 | 1,019 | 1,125 | 1,064 | 1,035 | 1,010 | 1,002 |
| Average (with) | 1,019 | 1,016 | 1,021 | 1,179 | 1,097 | 1,059 | 1,010 | 1,002 |
| MAPE (without) | 8,7% | 7,8% | 5,7% | 66,6% | 27,9% | 19,4% | 1,0% | 0,3% |
| MAPE (with) | 8,6% | 7,5% | 5,8% | 39,7% | 23,3% | 17,2% | 1,0% | 0,3% |

**Table 5: dosimetric impact of the 95% bladder outliers methodology**

Table 5 shows that the elimination of the bladder outliers, which induced an important gain on the bladder dosimetric error (MAPE). For the rectum and prostate, there was no change and the same errors were observed. In detail we found that the ratio of D50 (between first and second scan) is significantly decreased (from 66.6% to 39.7%), the bladder average dose significantly reduced and the bladder EUD ratio decreased (from 19.4% to 17.2%). Hence, the process of eliminating data outliers doesn't seem to affect the dose to the target volume and the prostate centroid displacement (0.3cm) is compensated by the PTV margins.

### 3.2.2 Elimination of rectum parameter Outliers

Using the same process as previously described, approximating the normal distribution of rectal volume ratio, we obtained an average volume ratio of 1.02 with a standard deviation of 0.205. We will quantify the impact of eliminating all rectum modifications with volume ratio outside [0.613 - 1.434].

|  | C/C | R/L | A/P | 3d | Vol (MAPE) |
|---|---|---|---|---|---|
| without | 0,47 | 0,28 | 0,31 | 0,63 | 17,69% |
| with | 0,45 | 0,29 | 0,3 | 0,61 | 16,70% |

Table 6: geometrical impact for rectum of the 95% rectum outliers methodology

As in the case for the bladder, Table 6 shows that this induces very little displacements.

|  | Ratio Rectum | | | Ratio Bladder | | | Ratio CTV2 | |
|---|---|---|---|---|---|---|---|---|
|  | D30 | Dmean | EUD | D50 | Dmean | EUD | D95 | Dmean |
| Average (without) | 1,016 | 1,015 | 1,019 | 1,125 | 1,064 | 1,035 | 1,010 | 1,002 |
| Average (with) | 1,002 | 1,006 | 1,010 | 1,143 | 1,076 | 1,045 | 1,010 | 1,002 |
| MAPE (without) | 8,7% | 7,8% | 5,7% | 66,6% | 27,9% | 19,4% | 1,0% | 0,3% |
| MAPE (with) | 8,0% | 7,4% | 5,2% | 67,2% | 27,6% | 19,1% | 1,0% | 0,3% |

**Table 7: dosimetric impact of the 95% rectum outliers methodology**

On Table 7, it is interesting to note that the outliers elimination method does not improve the mean dose ratio, D30, nor the EUD for the rectum. In fact, rectal dose modifications are not only related to the rectal volume (shift in the A / P direction), but also to the movements caused particularly by gas (local contraction and expansion). The simple ratio of volumes is not sufficient to establish a dosimetric rule. The absolute volume of the rectum may be more specific and could act as a predictive parameter.

### 3.2.3 Elimination of bladder and rectum parameter Outliers

In this case the largest bladder and rectum volume ratios were eliminated (5% of bladders and rectums).

|  | Ratio Rectum | | | Ratio Bladder | | | Ratio CTV2 | |
| --- | --- | --- | --- | --- | --- | --- | --- | --- |
|  | D30 | Dmean | EUD | D50 | Dmean | EUD | D95 | Dmean |
| Average (without) | 1,016 | 1,015 | 1,019 | 1,125 | 1,064 | 1,035 | 1,010 | 1,002 |
| Average (with) | 1,003 | 1,007 | 1,011 | 1,200 | 1,111 | 1,070 | 1,009 | 1,001 |
| MAPE (without) | 8,7% | 7,8% | 5,7% | 66,6% | 27,9% | 19,4% | 1,0% | 0,3% |
| MAPE (with) | 7,8% | 7,0% | 5,2% | 39,3% | 22,8% | 16,8% | 1,0% | 0,3% |

**Table 8: dosimetric impact of the 95% bladder and rectum outliers methodology**

Table 8 shows clearer results than for the two previous cases. Improvements are undoubtedly related to the bladder, with the D50 ratio (-27.3%) followed by the medium dose (-5.1%) and the EUD (-2.6%). The gain to the rectum is also interesting but less important, with about 1% gain for all parameters. The best method consists in repeating only 5% of the dosimetries for patients with a large change in bladder volume (in the present study, 5% of outliers represent a ratio superior to 250%). This process can reduce the risk of mismatch between the initial scan and those of the following days. The use of this process for the rectal volume ratio is less interesting, and is only a secondary correction.

### 3.2.4 A global solution

We have seen that by repeating 5% of the dosimetries (and 10% if we consider both bladder and rectum cases), it is possible to improve the quality of the radiotherapy. It is now of interest to find the link between the number of repeated dosimetries and the gain generated (D30, $D_{mean}$….). Firstly, we need to find the parameters, such as organ volume or isocentre shift, that are most highly correlated to the EUD ratio for the bladder and rectum. The selected parameters should be strictly orthogonal and not correlated with each other as they provide the basis for a linear regression.

### 3.2.4.1 The bladder case

To complete the statistical study, the EUD ratio needs to be transformed by values superior to 1, so that a ratio of 0.9 becomes 1.11 (= 1/0.9). This modification prevents any compensation of the errors. To determine the parameters having most influence on the bladder EUD, we used the classical correlation coefficients associated to a Student T-test. We obtained parameters which have a CC statistically not equal to zero, and so which are significant: the D50 bladder (CC = 0.84), the average dose to the bladder (CC = 0.97), the ratio of the bladder volume (CC = -0.7), the 3-D displacement of the bladder (CC = 0.77) and the 3-D movement of the prostate (CC = 0.5). Note that the average dose is more representative of the risk (EUD bladder) than the D50, which could allow us to change our methodology to validate the bladder dose (actually D50 < 62 Gy). None of the dosimetric parameters (average dose to the bladder, D50…) are interesting for the proposed study, and we instead chose the two parameters, volume ratio of the bladder and 3-D displacement of the prostate. We used the fact that the bladder centroid displacement is a linear combination of displacement of the prostate's isocentre and bladder volume. If the volume ratio is large then the isocentre moves, and if it is not properly positioned on the prostate, it may also cause a shift on the bladder.

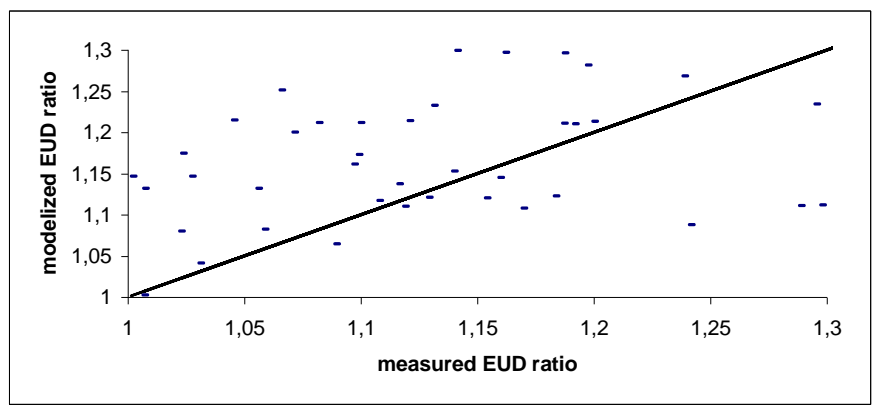

**Figure 6: Modelized and measured bladder EUD ratio (normalized Root Mean Square Error=12,5%). The line represents the equation y=x.**

We find a regression for bladder EUD (F-Fisher = 11.5 and Figure 6) as such:

$$EUD_{bladder} = 0.937 \ _{(p<0.001)} + 0.26 \times 3d\_prostate \ _{(p<0.001)} + 0.062 \times ratio\_bladder \ _{(p=0.04)}$$

If we consider the use of the daily CBCT for the treatment and a systematic focus on the prostate's isocentre (3d_prostate=0), the second term of this equation can be ignored as it is equal to zero. The objective is to say that if we fixed the problem of the isocentre (actual IGRT), we can only work on the volume of the bladder. The regression is limited to:

$$EUD_{bladder} = 0.937 + 0.062 \times ratio\_bladder$$

So the bladder volume ratio can be imposed by the difference on the EUD that we allow.

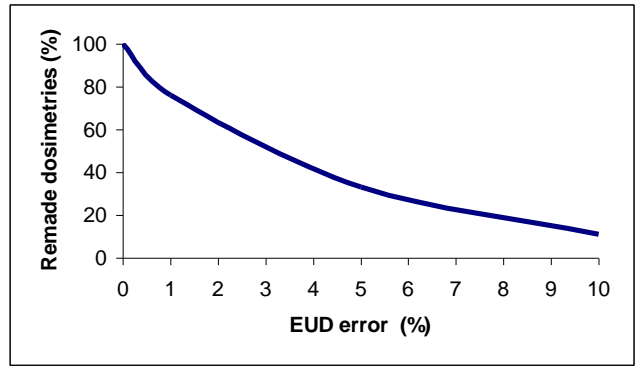

**Figure 7: Link between the number of dosimetry to remake, and the bladder EUD difference authorized**

On Figure 7, we can see that a threshold at 10% for the EUD (representing a tolerance of 10% for EUD compared to the initial value on the first dosimetry), represents an interval of between 2.63 and 0.38 for the bladder volume ratio, and corresponds to 11% of the misadjusted dosimetries. For a threshold at 5% (for bladder EUD error), there is an interval of between 0.54 and 1.8, thus 33% of dosimetries need to be remade, and for a threshold at 1%, there is a margin of 1.18 - 0.85 which corresponds to 76% of dosimetries to adapt.

3.2.4.2 The rectum case

Using the same methodology as previously described, we obtained for the rectal volume the parameters that most influence the rectum EUD, namely the D30 (CC = 0.85), the average dose (CC = 0.84), the volume ratio (CC = 0.57), the 3-D displacement (CC = 0.3) and the 3-D movement of the prostate (CC = 0.31).

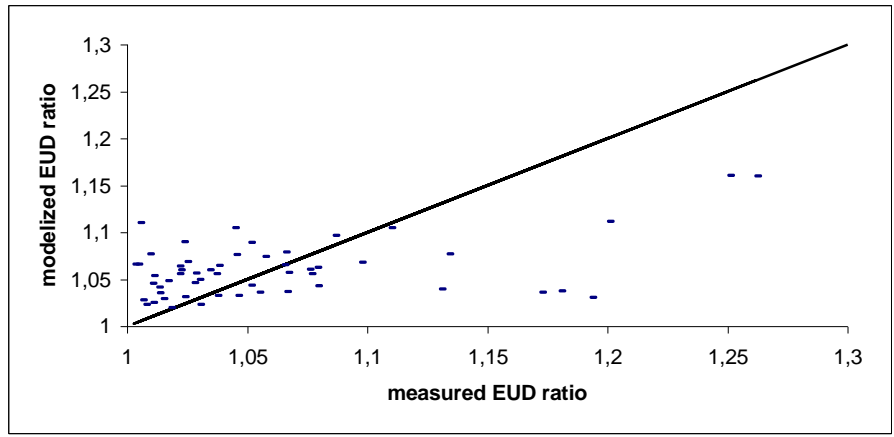

**Figure 8: Modelized and measured rectal EUD ratio (normalized Root Mean Square Error=5.1%). The line represents the equation y=x**

We get a regression (F = 7.97 and Figure 8) as follows:

$$EUD_{rectum} = 0.86\ _{(p<0.001)} + 0.05 \times 3d\_prostate\ _{(p=0.05)} + 0.144 \times ratio\_rectum\ _{(p=0.003)}$$

As in the previous case, the use of daily CBCT cancels the 3-D prostate coefficient, which tends to near 0 when the number of CBCT increases and the new relationship becomes:

$$EUD_{rectum} = 0.86 + 0.144 \times ratio\_rectm$$

Fig. 9 shows that for 10% of EUD ratio (tolerance of 10% for EUD compared to the initial value on the first dosimetry), there is an interval in the bladder volume ratio of between 1.67 and 0.6, which represents 4% of the dosimetries which need to be remade.

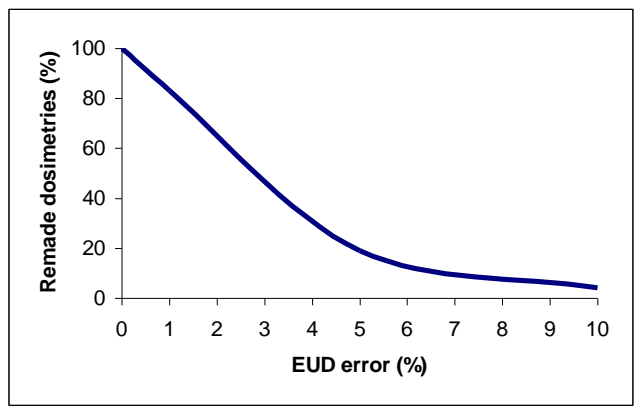

**Figure 9: Link between the number of dosimetry to remake, and the rectum EUD difference authorized**

For 5% of EUD ratio, there is an interval of between 1.32 and 0.76 (19% of dosimetries to remake), and for 1%, there is an interval of between 1.04 and 0.96 (83% of dosimetries to redo).

# 4 Conclusion and perspectives

Organ movement is an important aspect to consider in 21$^{th}$ century radiotherapy. Many authors have already studied this problem and have developed the IGRT process, which consists in positioning the beams in the target volume with a cone beam acquisition. This approach ensures that tumor control is maximal. In the proposed study, the approach is different as both the target volume and the organs at risk are studied, a methodology called the DGRT. The cone beam CT is necessary to better position the beam isocentre and the dosimetry is updated throughout the radiotherapy session. This method is based on bladder and rectal volume modifications, and if this variation is important, the dosimetry must be corrected. The future of radiotherapy must consider organ movements, and tools such as CBCT generate many images which need to be interpreted. Our methodology is dedicated to image data selection in order to eliminate all unnecessary manipulations. However, it will be necessary to complete the results shown in the present study (only 2 CT acquisitions for each patient) with studies including many CT acquisitions per patient. Some results like the cross-correlations will be certainly identical, but other like the relation between the EUD ratio and the volume ratio (see section 3.2.4), must be verified in this case.

## Acknowledgments

The authors thank the MV2A association members (M. de Mari, S. Cardi, V. Leandri, F. Savigny and C. Yousse) for their help during the CT acquisitions.

## Figures legends

Figure 1: Methodology for CT acquisitions
Figure 2. Displacement of the bladder centroid
Figure 3: Rectum centroid displacements
Figure 4: Prostate centroid displacement (cm). The line is the average, and the dotted line the average + 2$\sigma$
Figure 5 : CTV volume ratio between the two scans, line is the average and the dotted line the average +/- 2 $\sigma$ (standard deviation)
Figure 6: Modelised and measured bladder EUD ratio (normalized Root Mean Square Error=12,5%). The line represents the equation y = x
Figure 7: Link between the number of dosimetry to remake, and the bladder EUD difference authorized
Figure 8: Modelised and measured rectal EUD ratio (normalized Root Mean Square Error=5.1%). Line is the graph of y=x
Figure 9: Link between the number of dosimetry to remake, and the rectum EUD difference authorized

## Légendes des Figures

Figure 1: Methodologie pour les acquisitions CT
Figure 2. Déplacement de la véssie (centroide)
Figure 3: Déplacement du rectum (centroide)
Figure 4: Déplacement de la prostate (cm). Le trait plein représente la moyenne, et les pointillés la moyenne+2$\sigma$
Figure 5 : Ratio du volume du CTV entre les 2 scans. Le trait plein représente la moyenne, et les pointillés la moyenne +/- 2 $\sigma$ (écat type)
Figure 6: Ratio de l'EUD vésicale mesuré et modélisé (normalized Root Mean Square Error=12,5%). La ligne est la droite d'équation y=x
Figure 7: Relation entre le nombre de dosimétries à réévaluer et la différence de l'EUD vesicale souhaitée
Figure 8: Ratio de l'EUD rectale mesuré et modélisé (normalized Root Mean Square Error=12,5%). La ligne est la droite d'équation y=x
Figure 9: Relation entre le nombre de dosimétries à réévaluer et la différence de l'EUD rectale souhaitée

**Tables legends**

Table 1: absolute difference for the centroids (in cm), and the volumes between the first and second CT acquisition (*is the mean value and 95% confidence interval=mean±2.Std_dev.$(54)^{-1/2}$)
Table 2: correlation coefficients (CC) for the bladder, rectum and prostate parameters
Table 3: cross interactions between prostate and bladder and rectum parameters (top), and cross correlation between bladder and rectum parameters (bottom)
Table 4: geometrical impact for bladder of the 95% bladder outliers methodology
Table 5: dosimetric impact of the 95% bladder outliers methodology
Table 6: geometrical impact for rectum of the 95% rectum outliers methodology
Table 7: dosimetric impact of the 95% rectum outliers methodology
Table 8: dosimetric impact of the 95% bladder and rectum outliers methodology

**Légende des tableaux**

Table 1: différence absolue pour les isocentres (cm), et différence entre les volumes de la première et de la seconde acquisition CT (* représente la moyenne suivi de l'intervalle de confiance 95% )
Table 2 : coefficients de corrélation (CC) pour les paramètres de la vessie, du rectum et de la prostate
Table 3 : interactions croisées entre les paramètres de la prostate, et ceux de la vessie et du rectum (en haut), ainsi que les corrélations croisées entre les paramètres du rectum et de la vessie (en bas)
Table 4 : impact géométrique pour la vessie de la méthodologie des « 95% bladder outliers »
Table 5 : impact dosimétrique de la méthodologie  « 95% bladder outliers »
Table 6 : impact géométrique pour le rectum de la méthodologie des « 95% rectum outliers »
Table 7 : impact dosimétrique de la méthodologie  « 95% rectum outliers »
Table 8 : impact dosimétrique de la méthodologie  « 95% rectum and bladder outliers »